\documentclass{ws-p8-50x6-00}
\renewcommand{\rightnote}{}

\begin{document}

\title{Vector spectrum and color screening in two color QCD 
at nonzero $T$ and $\mu$}

\author{Bartolom\'e ~All\'es }
\address{Dipartimento 
           di Fisica, Universit\`a di Milano--Bicocca and INFN 
           Sezione di Milano, I-20126 Milano, Italy}
\author {Massimo~D'Elia} 
\address{Dipartimento di Fisica, 
           Universit\`a di Genova and INFN Sezione di Genova, 
           I-16146 Genova, Italy}
\author {Maria- Paola Lombardo }
\address{INFN Sezione di Padova, and Laboratori di Frascati, I-00044 
Frascati, Italy}
\author {Michele Pepe }
\address{Laboratoire de Physique Th\'eorique, 
                          Universit\'e de Paris XI, 91405 Orsay Cedex, France
and ITP, University of Bern, CH-3012 Bern, Switzerland}

\maketitle

\abstracts
{We discuss a few aspects of the phase diagram of
two color lattice QCD: we investigate the
long distance screening  analyzing the behavior of
the interquark potential at large distances; 
we present a first set of results for vector 
mesons and diquarks; we note similarities and differences
between  features at high temperature
and high baryon density. }

\section {Introduction}

In this note we report on a numerical investigation of the
phase diagram of two color QCD in the temperature--chemical potential
plane. In this theory, the fermion determinant is positive at nonzero
baryon density~\cite{diqcon,HKLM} and lattice simulations are possible.

Several studies of this model have already 
appeared~\cite{modellattice,temp,vecon} and,
in particular, the phase diagram in the temperature--density plane at nonzero
bare quark mass has been studied in
some detail~\cite{ph4f,ect,kim}. 
There are three different phases: a hadronic phase
with confinement and spontaneous (approximate) 
chiral symmetry breaking, a superfluid phase
with diquark condensation and a plasma phase. For a nonzero quark
mass and a zero diquark source, which is our case, 
the superfluid line is a bonafide phase transition, the
other lines are either crossover or first order transitions. 
It has been shown~\cite{ph4f,kim} 
that, for some values of $\mu$, either  
$\langle \bar\psi \psi\rangle$ or
$\langle \psi \psi\rangle$
 are non--monotonic function of temperature.
An earlier, sketchy proposal of the phase
diagram~\cite{ect}  suggests a non--monotonic behavior
for the diquark condensate.  Such behaviors are rather common 
in condensed matter (e.g. superconductors) but unusual in particle
physics: that adds to the interest of two color QCD.

Our aim is to characterize the various phases and their 
associated phenomena
by continuing the study of the spectrum of mesons and diquarks and by
exploring in detail the behavior of purely gluonic observables.

Here we will discuss only a few aspects of our study: the behavior of
the Polyakov loop and its correlators, and a first set of results
for the vector meson and diquark spectrum. A complete presentation, 
which will include topological~\cite{topol} and thermodynamics results,
will appear in a future publication~\cite{new}.

\section{The simulations}

Numerical simulations of two color QCD with eight continuum
flavors have been carried out on a $14^3 \times 6$
lattice by an HMD algorithm~\cite{HKLM}, for several values of
temperatures, masses and chemical potentials.
We have considered chemical potentials between zero and the lattice
saturation and temperatures ranging from $T \simeq 0$ to $T> T_c$.

Typically, we have run for O(10000) MD steps, with dt = .02 and taking
measures every 40 steps.

A few test runs were performed with dt = .04 and dt = .08 in order to check
that discretization errors are small at dt = .02. 
The accuracy of the inversion of the Dirac operator has been tested 
by use of the lattice Ward identity.

\section{Color screening}

It is interesting to remark 
that for two color QCD the Polyakov loop $P$ is a real quantity: hence,
  the Polyakov loop correlator
defining the strength of the quark-quark
interaction $ \langle P P (R) \rangle / \langle P \rangle^2$, 
is the same as the quark--antiquark one
$ \langle P P^\dagger (R) \rangle / |\langle P\rangle|^2$ .
This property  remains true also
at $\mu \ne 0$.

Note however, that this is the only special feature of the Polyakov
loop and its correlators in two color QCD: all of the discussion can be 
carried out for a generic number of color $N_c$. 
For this reason, it might well
be that some of the finding presented here carry over to real QCD.

The confined phase in the quenched approximation is characterized by
an interquark static potential which rises linearly at any distance
$\lim_{R \to \infty} V(R) \propto \sigma R$,  $\sigma$ being
the string tension.

When quark degrees of freedom start to be dynamical a new phenomenon
takes place. The color flux tube between two static quarks can
now break by pair production of dynamical quarks popping out of the
vacuum and screening the two static sources (string breaking): $Q\bar Q
\rightarrow (Q\bar q)(q\bar Q)$. The potential does not rise
indefinitely anymore but flattens out at sufficiently large distance
$R_0$.

String breaking in the hadronic phase has been 
observed at a moderately high temperature~\cite{DKKL},
and it has also been
suggested that the real particles present in a dense system
further favors the breaking of the string~\cite{EKKL}.

In the following, we will study the behaviour of the interquark potential
as a function of $\mu$, and we will contrast its behaviour with that
observed as a function of temperature.

\begin{figure}
\epsfxsize=25pc 
\epsfbox{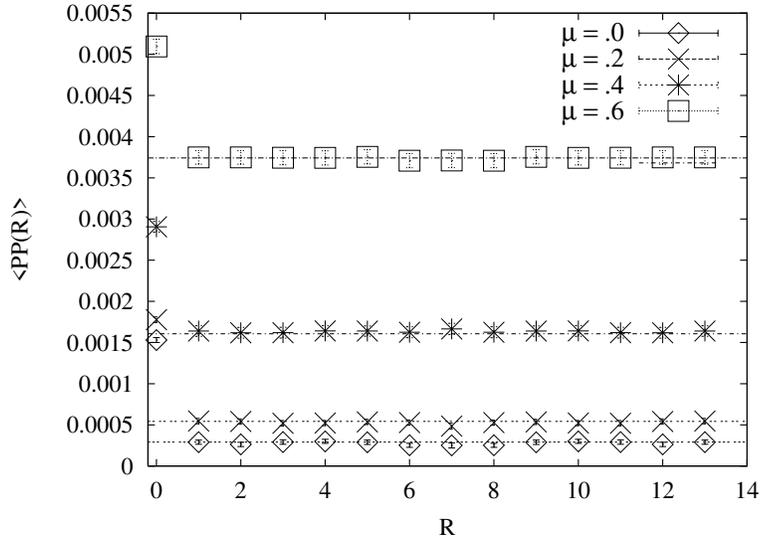}
\caption{Polyakov loop correlation for $\beta = 1.5$ and
various  chemical potentials:
screening and string breaking grow rapidly around $\mu_c$.
The horizontal lines are drawn in correspondence of $\langle P\rangle ^2$, demonstrating
the cluster property of the correlators.}
\label{fig:polcor}
\end{figure}
\begin{figure}
\epsfxsize=14pc 
\epsfysize=12pc 
\epsfbox{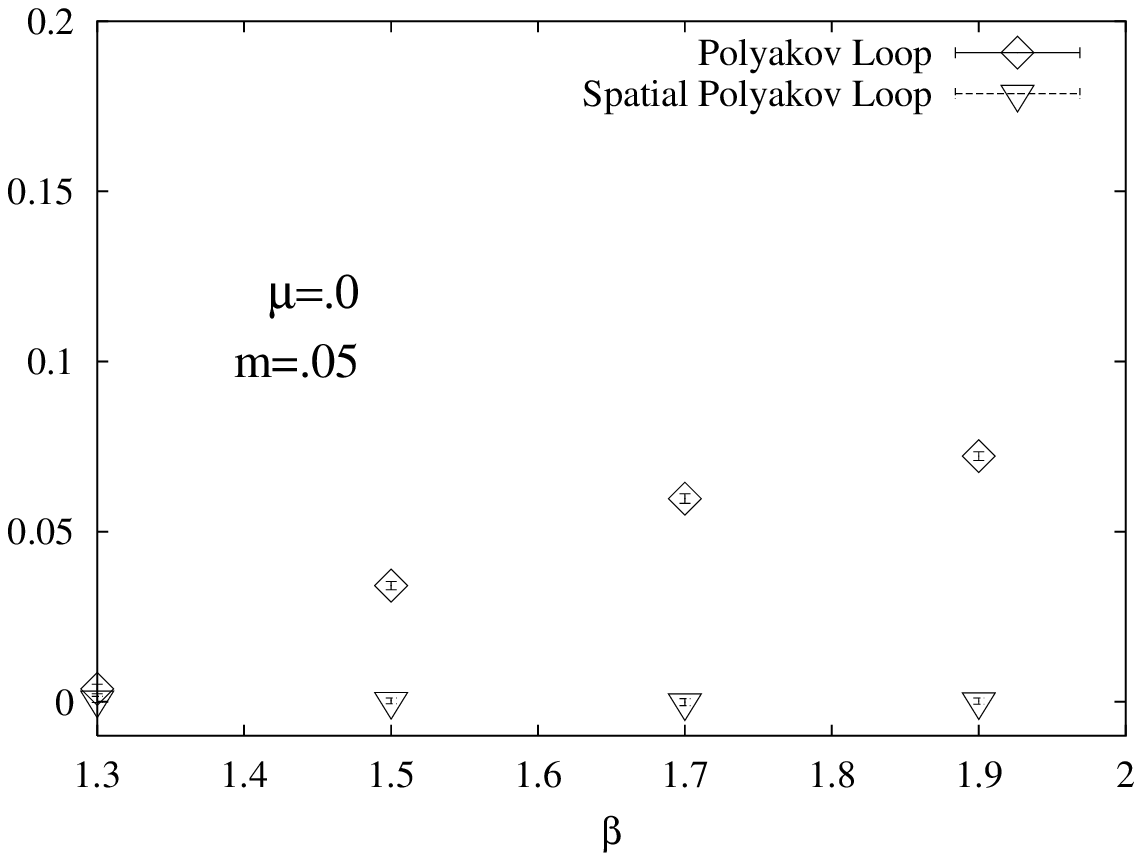}
\epsfxsize=14pc 
\epsfysize=12pc 
\epsfbox{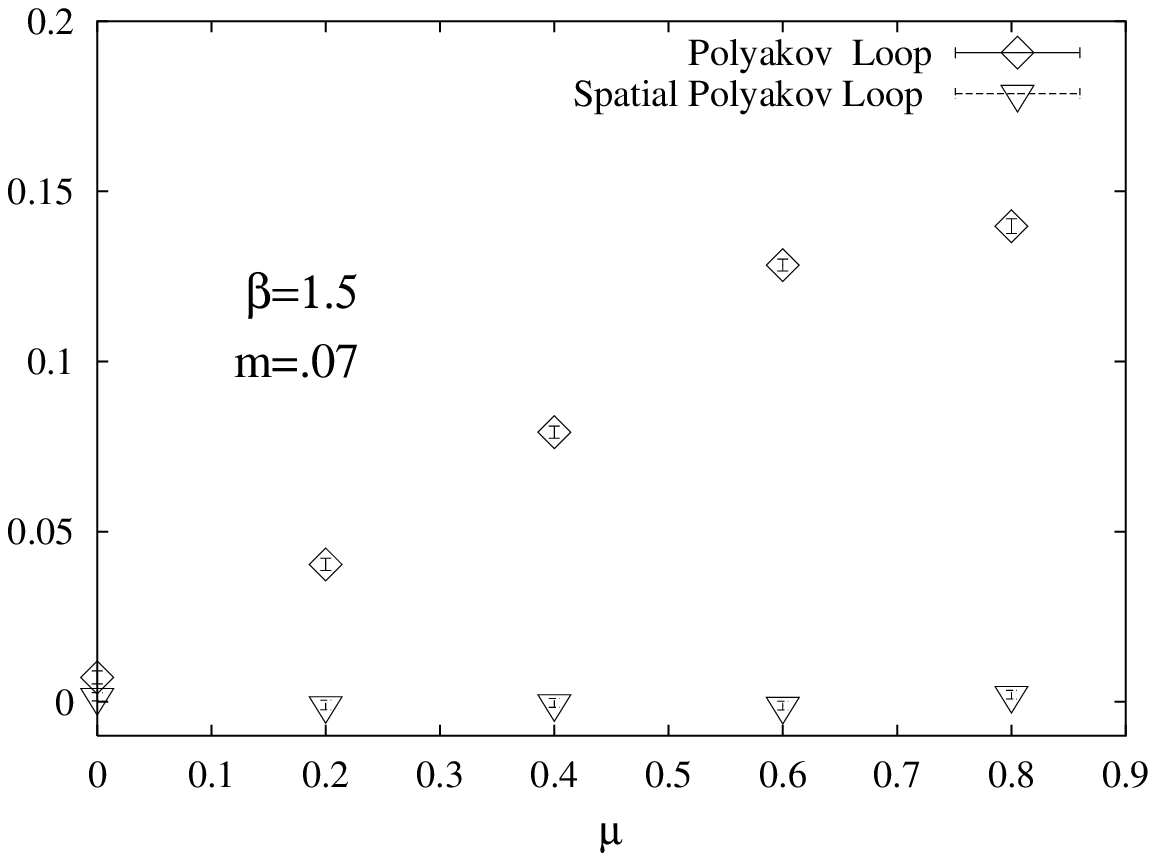}
\label{fig:poltot}
\caption{Polyakov loop, and spatial Polyakov loop, as a function of
$\beta$ (left) and $\mu$}
\end{figure}
Consider  the Polyakov loop at spatial coordinates $\vec n = (n_1,n_2,n_3)$
\begin{equation}
P(\vec n) = (1/N_c) Tr \prod^{N_\tau -1}_{\tau = 0} U_0(\vec n, \tau) \; .
\end{equation}
We average $P(\vec n)$ on spatial planes to build:
\begin{equation}
 P_3 (n_3) = (1/{N_s}^2) \sum_{n_1,n_2 = 1}^{N_s} P(n_1,n_2,n_3) \; 
\end{equation}
$P_1$ and $P_2$ are similarly defined.

By use of $P_i (n_i)$ we finally arrive at
the zero-th momentum projected
correlation function of the Polyakov line
\begin{equation}
\langle PP^\dagger(R)\rangle = \frac{1}{3 N_s} \sum_{i=1}^{3} \sum_{n_i=1}^{N_s} \langle  P_i(n_i)
 P_i^\dagger (n_i + R) \rangle \; .
\label{eq:polc}
\end{equation}

\begin{figure}
\epsfxsize=24pc 
\epsfysize=15pc 
\epsfbox{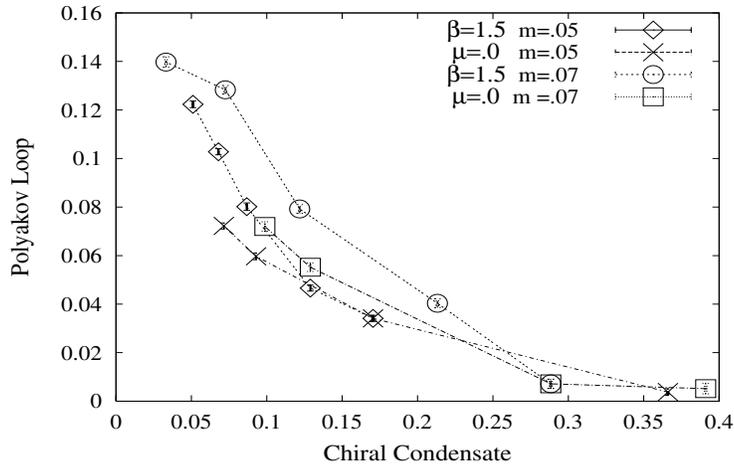}
\caption{Polyakov loop versus the chiral condensate}
\label{fig:ptvspbp}
\end{figure}
\begin{figure}
\epsfxsize=14pc 
\epsfysize=12pc
\epsfbox{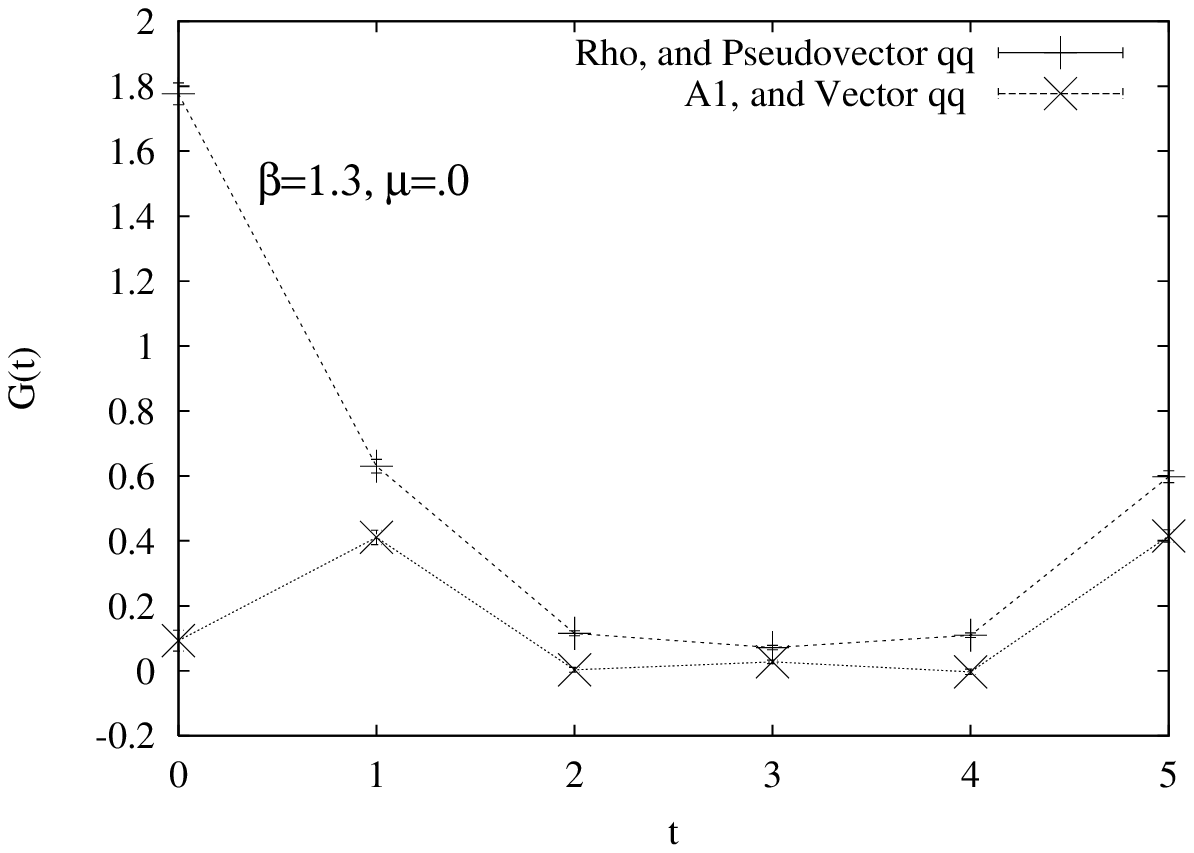}
\epsfxsize=14pc 
\epsfysize=12pc
\epsfbox{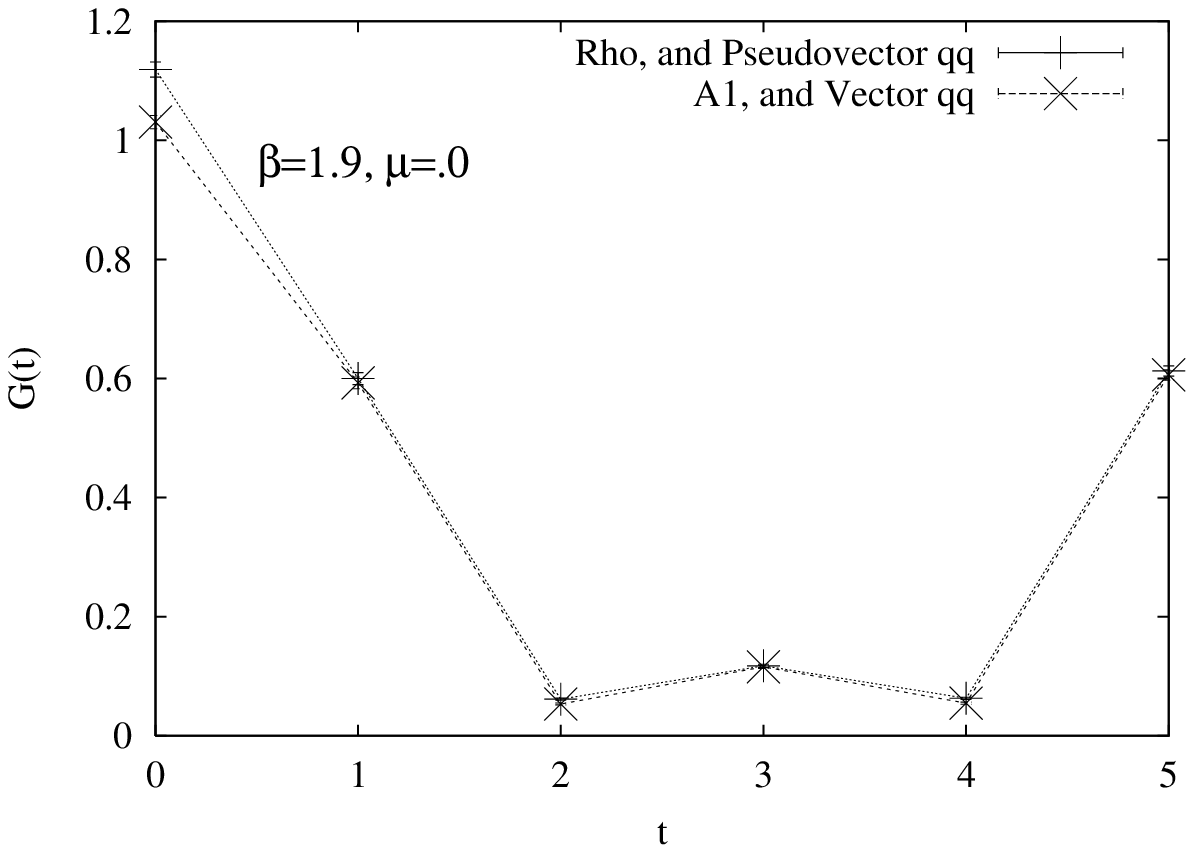}
\\
\noindent
\epsfxsize=14pc 
\epsfysize=12pc
\noindent 
\epsfbox{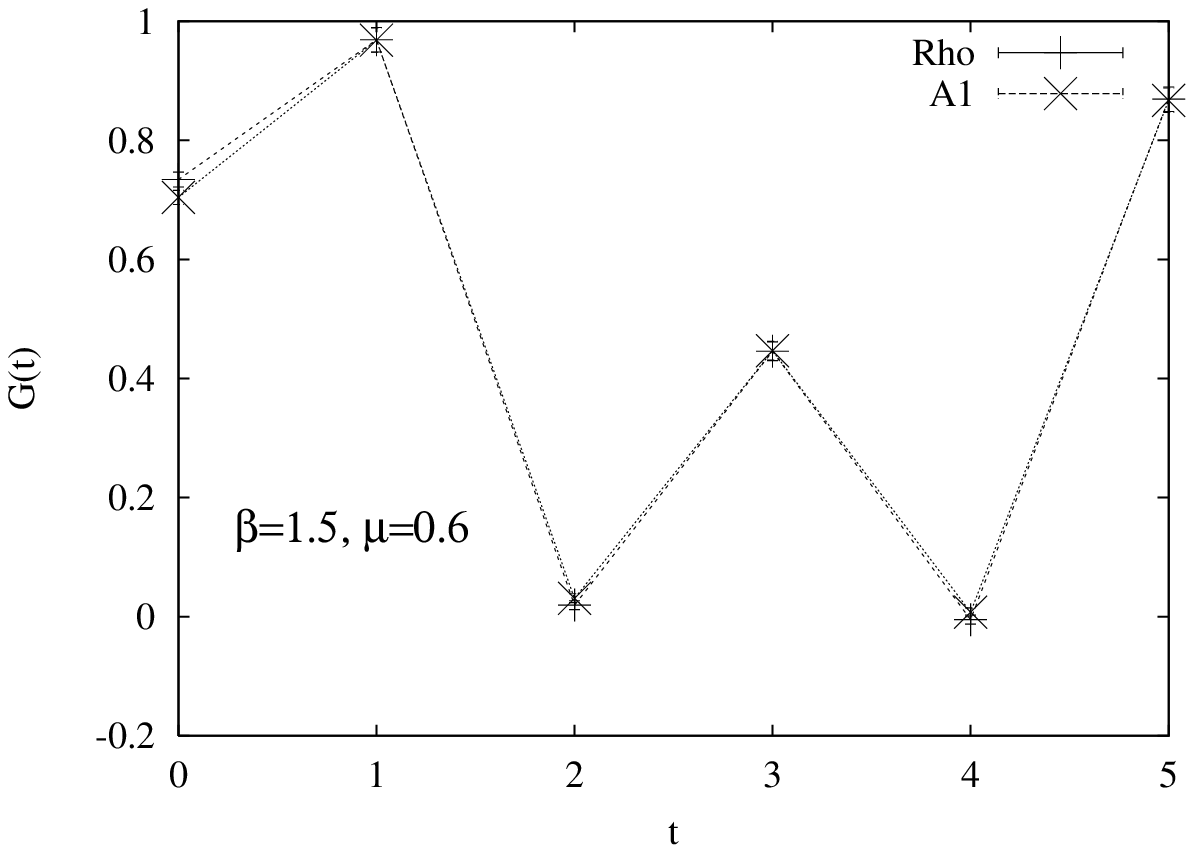}
\epsfxsize=14pc 
\epsfysize=12pc 
\noindent
\epsfbox{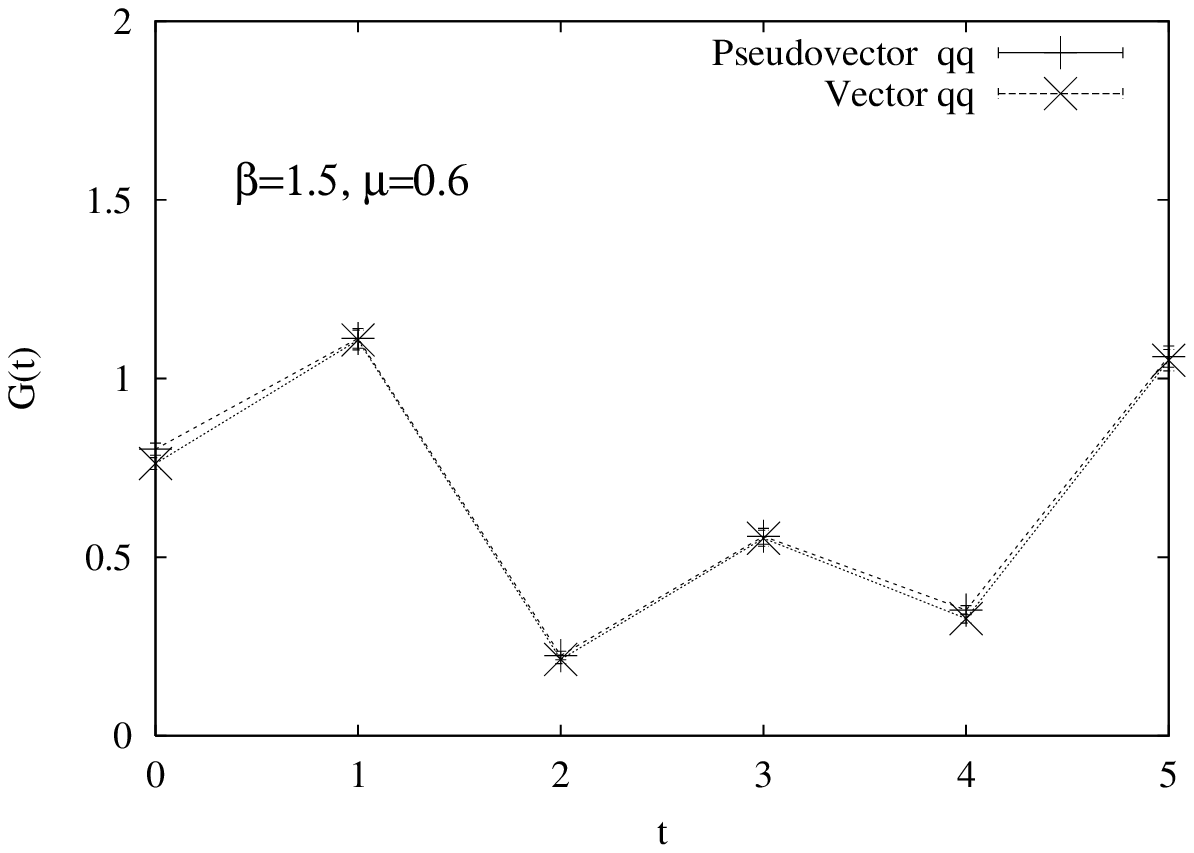}
\caption{Vector mesons and diquarks propagators $G(t)$ as a function of 
Euclidean time $t$. Upper, left: vector and pseudovector meson propagators
in the hadronic phases. Vector diquarks are degenerate. Upper, right:
as in left part of the diagram, but in the plasma phase: chiral
symmetry is restored. Lower, left: vector mesons at high density; 
the propagators are degenerate, as it happens at high T, but chiral
symmetry is still broken. Note anomalous spectral behaviour.
Lower, right: vector diquarks at high
density; time reversal symmetry is broken, and we note possible hints of
vector condensation}
\label{fig:vettori}
\end{figure}
In Fig.~\ref{fig:polcor} we show the results for the Polyakov loop
correlators at various values of the chemical potential.
At high chemical potential we observe the expected
enhanced screening, since the correlation
functions approach a constant value at large distance.
The gap between
the plateaux at $\mu=0.2$ and $\mu=0.4$ in Fig.~\ref{fig:polcor}  
 suggests  the passage to a deconfined phase. We have
then a direct evidence of the effect of the chemical potential
on the gauge fields.

We note that the  trend with chemical potential closely
resembles that observed with temperature 
\cite{DKKL}: in both cases 
we have signals of deconfinement. 

In Fig.~2 the dependence of the Polyakov 
loop  on $\mu$ (left hand side) and on $T$
(right hand side) is contrasted with that of the spatial 
Polyakov loop, superimposed in the same plots. 
Again we note the similarity between
the behaviour 
at high temperature and that
at high density : in both cases, only the temporal
loop is affected.

These results provide a
direct observation
of long range screening in a dense gauge system, obtained from
a first principle study of the model, further confirming 
other observations\cite{temp}. They are thus providing an
important confirmation of the standard expectation of the structure of a 
dense medium.

A more detailed analysis can give us more information on the
screening properties. We note that some theoretical guidance can also be
offered by low energy effective string models~\cite {GP}, 
which provide the following expression (simplified for the sake of the
present discussion) for the point-to-point correlators:
\begin{equation}
\langle PP^\dagger(R)\rangle = \alpha e^{- \sigma_0 R N_t} + 
(N_cN_f)^2 \beta e^{- 2 \sigma_0 R_0 N_t} \; .
\end{equation}

The above equation shows the
 expected exponential dependence on the breaking length
$R_0$, hence on the mass scale, and we also note the dynamical factors $\alpha$
and $\beta$. It is then reasonable to assume that they depend on 
temperature and density: we will address this point in our 
future publication.

Finally,  in Fig.~\ref{fig:ptvspbp}
 we  study the correlation between  $|\langle P\rangle|$
and $\langle\bar \psi \psi\rangle$ by plotting one versus the other. 
We plot there two sets of data: firstly, those obtained at fixed $\beta =1.5$,
and variable $\mu$. Secondly, those at $\mu=0$, and variable $\beta$.
We see
that  chiral condensate and Polyakov loop are well correlated, 
for both data sets , and that the Polyakov loop looks consistently larger for
the first data set $, \beta =1.5, \mu \ne 0$, supporting the
idea that a finite density of real particles further favors
string breaking~\cite{EKKL}.

\section{The vector sector}

The vector meson ($\bar q \gamma_i q$) and diquark 
($q \gamma_i q$) and antidiquark ($\bar q \gamma_i \bar q $) 
propagators at 
$\mu=0$ are constructed from the quark propagator,
as explained in early work~\cite{HKLM} for the scalar spectrum, 
modulo the inclusion of the appropriate
$\gamma$ matrix (or, rather, of its staggered fermion representation).

From  the Pauli--Gursey symmetry of the SU(2) Action, 
the exact degeneracy of
the vector (V) meson propagator with the pseudovector (PV) diquark
propagator follows,
as well as that of the pseudovector meson propagator with the vector 
diquark propagator. Analogous degeneracies have been noted in the scalar
sector \cite{HKLM}.

Finite density spectroscopy analysis 
has been already introduced and discussed~\cite{KLS0}. 
The following symmetries should hold
true in the ensemble:
\begin{eqnarray}
G(t) & = & G(T-t) \qquad \mbox{\rm for mesons} \\
G(t,\mu) & = & G(T-t, -\mu) \qquad\mbox{\rm for diquarks}
\end{eqnarray}
In addition to that, remember that staggered mesons contain states
with different lattice parity (the `oscillating' channel):
for instance, the vector meson contains both the $\rho$ and the B particle,
while the pseudovector contains the A1 and again the $\rho$:
\begin{eqnarray}
G_V(t) &=& a(e^{(-m_\rho t)} + e^{m_\rho(T-t)}) 
+ (-1)^tb(e^{(-m_B t)} + e^{m_B(T-t)}) \\
G_{PV}(t) &=& a(e^{(-m_{A1} t)} + e^{m_{A1}(T-t)}) 
+ (-1)^tb(e^{(-m_\rho t)} + e^{m_\rho(T-t)})
\end{eqnarray}

Since mesons do not `feel'
the effects of the chemical potential the masses of the forward 
and backward propagating meson states are identical.

For $\mu\neq 0$ we expect different
 forward and backward masses in the diquark channels,
reflecting the different backward and forward propagations in the 
dense medium. The occurrence of condensation, in either channel,
should be signaled by the appearance of a constant term at
large distance.

Let us look now at the diagrams, Fig.\ref{fig:vettori}: the upper left diagram
shows the vector and pseudovector meson propagators at low temperature,
zero chemical potential. We have verified
that they are degenerate with the vector diquarks of opposite parity.
By increasing the temperature  we observe chiral symmetry restoration 
in the vector channel (Fig.\ref{fig:vettori}, upper
right diagram): all of the four vector states (mesons and 
diquarks) are now degenerate.
At nonzero density mesons and diquarks with opposite parity
are no longer degenerate, 
as anticipated. Let us follow independently mesons
and diquarks then. 

In  
Fig.\ref{fig:vettori} ,  left lower diagram, we see the
vector and pseudovector mesons for $\mu > \mu_c$: 
they are degenerate, as it happens at high T: of course
this does not signal chiral symmetry restoration , since
chiral symmetry is still broken by the diquark condensate. Simply, it
tells us that the chiral condensate, which is responsible for the
difference between these two propagators, is now zero.
This is the same which happened with scalars.
We observe also an important difference : the oscillating componenent
is increased, as if the B particle 
contribution were more important at high
density. This might indicate a significant difference of the spectral functions
at high temperature and density, suggesting that the spectral concentration
of the $\rho$ decreases at high density in comparison to that of the 
B meson.

One last remark concerns the behavior  of vector diquarks: in addition
to the expected asymmetry, we observe 
(Fig.\ref{fig:vettori} ,  right lower diagram) a behaviour
which is not inconsistent with a possible plateau. Obviously, also
a very low mass can mimick this behaviour, and
all the usual caveat apply  to this observation : first of all we should
study colder lattices, where condensation phenomena are more prominent.
Secondly, we should check finite size scaling. 
If this trend were to persist for a range
of lattice volumes then it  could be interpreted as evidence
for the predicted~\cite{vecon} vector condensation.

\section{Summary}

We have observed enhanced screening and string breaking in the
critical region for chiral symmetry. We have shown that chiral condensate
and screening properties are correlated both at finite temperature
and finite density. These observations suggest that the (pseudo)critical
line for chiral condensate and confinement run close to each other
in the phase diagram of two color QCD. Moreover we noticed that,
at comparable values of chiral condensate, string breaking is
slightly more pronounced at finite density, and low temperature,
 than at zero density and finite temperature. This supports the view
that recombination with real quarks in a dense system favors 
deconfinement~\cite{EKKL}.

We have observed exact degeneracies in the vector sector
of the spectrum. We have studied the pattern of chiral symmetry 
for vector mesons and diquarks at finite temperature
and density. Our results show a peculiar
behaviour in the meson vector sector at high density, noticeably different
from that at high temperature: spectral concentration of the B particle
seems dominating over the $\rho$, while at high temperature the opposite
holds true. Finally we have noted possible hints of vector condensation in
the diquark sector.

One final remarks concerns the possibility to use two color QCD to
test methods for simulating QCD at nonzero density. Some experiments
with the Glasgow
reweighting have already been performed~\cite {crompton}, while test of 
the analytic continuation from  imaginary chemical potential~\cite{immu}
are in progress~\cite{testsu2}.

\section*{Acknowledgments}

We wish to thank 
J.T. Lenaghan, F. Sannino, 
W. Sch\"afer,
K. Splittorff and
D. Toublan   for interesting conversations. 
This work has been partially supported by MIUR. M.P. thanks the 
European Union Human Potential Program (contract HPRN-CT-2000-00145, 
Lattice QCD).

\end{document}